\documentclass[a4paper,12pt]{article}
\usepackage{amsmath,amsfonts,amssymb,cite}

\newcommand{\be}{\begin{equation}}
\newcommand{\ee}{\end{equation}}

\begin{document}

\renewcommand*{\thefootnote}{\fnsymbol{footnote}}

\begin{center}
{\large\bf $W$-boson mass anomaly as a manifestation of spontaneously broken
additional $SU(2)$ global symmetry on a new fundamental scale}
\end{center}
\bigskip
\begin{center}
{\large S. S. Afonin\footnote{E-mail: \texttt{s.afonin@spbu.ru}}}
\end{center}

\renewcommand*{\thefootnote}{\arabic{footnote}}
\setcounter{footnote}{0}

\begin{center}
{\it Saint Petersburg State University, 7/9 Universitetskaya nab.,
St.Petersburg, 199034, Russia}
\end{center}

\begin{abstract}
Recently the CDF Collaboration has announced a new precise measurement of the $W$-boson mass $M_W$ that deviates from the
Standard Model (SM) prediction by $7\sigma$. The discrepancy in $M_W$ is about $\Delta_W\simeq70$~MeV and probably
caused by a beyond the SM physics.
Within a certain scenario of extension of the SM, we obtain the relation
$\Delta_W\simeq\frac{3\alpha}{8\pi}M_W\simeq70$~MeV, where $\alpha$ is the electromagnetic fine structure constant.
The main conjecture is the appearance of longitudinal components of the $W$-bosons as the Goldstone bosons of a
spontaneously broken additional $SU(2)$ global symmetry at distances much smaller than the electroweak symmetry breaking
scale $r_\text{\tiny EWSB}$. We argue that within this scenario, the masses
of charged Higgs scalars can get an electromagnetic radiative contribution which enhances the observed value of $M_{W^\pm}$
with respect to the usual SM prediction. Our relation for $\Delta_W$ follows from the known one-loop result for the
corresponding effective Coleman-Weinberg potential in combination with the Weinberg sum rules.


\end{abstract}

\bigskip

The CDF Collaboration at Tevatron has recently reported a new precise measurement of the $W$-boson mass that shows about $7\sigma$
deviation from the prediction of the Standard Model (SM)~\cite{CDF}. The newly discovered $W$-boson mass anomaly caused much
excitement among the specialists in Beyond the SM (BSM) physics since it is widely believed that the given discrepancy,
if confirmed in future experiments, is related with some new BSM physics.

The new measurement of the $W$-boson mass announced by the CDF Collaboration is~\cite{CDF}:
$M_W^\text{\tiny (CDF)}=80.4335\pm0.0094~\text{GeV}$. After combining with previous Tevatron measurement of $M_W$,
the following final Tevatron result was reported~\cite{CDF},
\be
\label{1}
M_W^\text{\tiny (Tevat)}=80.4274\pm0.0089~\text{GeV}.
\ee
This value exceeds the SM expectation~\cite{pdg},
\be
\label{2}
M_W^\text{\tiny (SM)}=80.357\pm0.006~\text{GeV},
\ee
by
\be
\label{3}
\Delta_W=70\pm11~\text{MeV}.
\ee
The result~\eqref{1} can be also combined with other previous measurements of $M_W$ by LEP2, LHC and LHCb experiments,
the SM prediction~\eqref{2} may be updated as well. All these variations are able to change the estimate of
discrepancy~\eqref{3} at the level of 10\% (for instance, the updated central values obtained in the global fit of
Ref.~\cite{deBlas:2022hdk} are $M_W^\text{\tiny (exp)}=80.413~\text{GeV}$ and $M_W^\text{\tiny (SM)}=80.350~\text{GeV}$,
see also Ref.~\cite{Lu:2022bgw}). It is seen thus that the anomaly in the $W$-boson mass is certainly present.
A more convincing argumentation is given in the original paper~\cite{CDF}.

Not surprisingly, the very recent publication by the CDF Collaboration has already caused an avalanche of theoretical papers
explaining the observed $W$-boson mass anomaly with the aid of some tantalizing new BSM physics
(see, e.g.,~\cite{Bhaskar:2022vgk,raby,Dcruz:2022dao,Appelquist:2022qgl,Evans:2022dgq} and numerous references therein).
Most of proposals seems to be centered around the idea of introducing additional fundamental scalar particles, typically a new
multiplet of Higgs bosons, which can contribute to the $W$-boson mass.

We will try to approach the problem partly against this mainstream. Our basic observation is that the magnitude of mass
anomaly~\eqref{3}, $\Delta_W\simeq0.001M_W$, is of the order of a typical first quantum correction in QED, i.e.,
of the order of $\mathcal{O}(\alpha/\pi)$, where $\alpha\approx1/137$ is the fine structure constant (for example, the famous
anomalous magnetic moment of electron, in the first approximation, is $a_e=\frac{\alpha}{2\pi}\approx0.001$). This observation
suggests that $\Delta_W$ may have mainly electromagnetic origin and the given electromagnetic correction was missed in
the previous SM predictions. Then the question is how this electromagnetic contribution arises? In the given Letter, we propose
a possible mechanism that leads to the quantitative prediction~\eqref{3}.

We will consider the electromagnetic correction $\Delta_W$ as an effect arising at distances less (possibly, much less) than
the Electroweak Symmetry Breaking (EWSB) scale, $r_\text{\tiny EWSB}\simeq (246~\text{GeV})^{-1}$, due to certain BSM physics
to be guessed. Our working option for this BSM physics at distances $r\ll r_\text{\tiny EWSB}$ will be the following:
Along with the standard $SU(2)_L$ gauge symmetry acting on the triplet of gauge bosons $(W^+,W^-,W^0)$ there exists
an additional $SU(2)'$ global symmetry acting on the same triplet of gauge bosons. For derivation of our result, however,
it will be convenient to regard $SU(2)'$ as a gauge symmetry acting on the second triplet of gauge bosons $(W'^+,W'^-,W'^0)$
and take the degeneracy limit at the end.
We suppose further that the triplet of Higgs scalars $(\phi^+,\phi^-,\phi^0)$ which is eaten by $(W^+,W^-,W^0)$
on the scale $r_\text{\tiny EWSB}$ due to the Higgs mechanism, on a ''truly fundamental'' level,
represents simultaneously the triplet of Goldstone bosons of spontaneously broken $SU(2)'$ part
of fundamental symmetry.

Within this scenario, we suggest that the charged scalars $\phi^+$ and $\phi^-$ can get an electromagnetic
contribution to the mass via the radiative corrections, $\Delta M_\phi = M_{\phi^\pm} - M_{\phi^0}>0$. This mass difference remains
at larger distances, $r\gtrsim r_\text{\tiny EWSB}$, and, via the Higgs mechanism, eventually translates into
\be
\label{4}
\Delta M_\phi=\Delta M_W=M_{W^\pm} - M_{W^0}.
\ee
The given effect, not taken into account in the SM quantitative predictions, leads then to the observed mass
anomaly~\eqref{3},
\be
\label{5}
\Delta_W=\Delta M_W,
\ee
which seems to be unaffected by the mixing of $W^0$ with the $B$-boson of $U(1)_Y$ gauge part in the SM.


Consider the two-point correlation functions of vector currents coupled to the $W$ and $W'$ bosons,
\be
\label{6}
\langle J_V^\mu J_V^\nu \rangle = (q^2\eta^{\mu\nu}-q^\mu q^\nu)\Pi_V(q^2), \qquad V=W,W'.
\ee
The difference of correlators $(\Pi_{W'}-\Pi_{W})$ represents an order parameter for the assumed spontaneous 
symmetry breaking. At large Euclidean momenta $Q^2=-q^2$, one can write the standard Operator Product Expansion (OPE) for $\Pi_V(Q^2)$.
In the field theories based on vector interactions with (initially) massless fermions, it is natural to
expect that the first contribution to $\left(\Pi_{W'}(Q^2)-\Pi_{W}(Q^2)\right)$ arises from four-fermion operators. The case of
spontaneous CSB in massless QCD represents a canonical example~\cite{svz}. Since the four-fermion operators have the mass dimension~6,
the OPE leads then to the behavior
\be
\label{7}
\left(\Pi_{W'}(Q^2)-\Pi_{W}(Q^2)\right)_{Q^2\to\infty}\sim\frac{1}{Q^6}.
\ee
The validity of~\eqref{7} will be crucial for our scheme.

Next we apply the method of Weinberg sum rules~\cite{Wein}. This method is based on saturation of correlators by a narrow resonance
contribution plus perturbative continuum equal for both correlators. Omitting the irrelevant subtraction constant, the Weinberg ansatz is
\be
\label{8}
\Pi_{W}(Q^2)=\frac{F_W^2}{Q^2+M_W^2}+\text{Continuum},
\ee
\be
\label{9}
\Pi_{W'}(Q^2)=\frac{F_{W'}^2}{Q^2+M_{W'}^2}+\frac{F_\phi^2}{Q^2}+\text{Continuum}.
\ee
The corresponding decay constants in residues are defined by
\be
\label{10}
\langle 0| J_V^\mu |V \rangle = F_V M_V \epsilon^\mu, \qquad V=W,W',
\ee
\be
\label{11}
\langle 0| J_{W'}^\mu |\phi \rangle = iq^\mu F_\phi.
\ee
Here $\epsilon^\mu$ denotes the polarization vector and $\phi$ is the triplet of Goldstone Higgs bosons of spontaneously broken
$SU(2)'$ symmetry. The parametrization~\eqref{11} emerges by virtue of the Goldstone theorem. Substituting~\eqref{8} and~\eqref{9}
into~\eqref{7} we get the relations
\be
\label{12}
F_{W}^2-F_{W'}^2=F_\phi^2, \qquad M_{W}F_{W}=M_{W'}F_{W'}.
\ee
The relations~\eqref{12} are in one-to-one correspondence with the old Weinberg sum rules~\cite{Wein}, in which the vector $\rho$, axial $a_1$
and pseudoscalar $\pi$ mesons play the role of $W$, $W'$ and $\phi$, correspondingly.

Initially, the Goldstone bosons $\phi^\pm$ and $\phi^0$ are degenerate in mass. But one can expect that the photon loops will generate
a potential, hence, an electromagnetic mass term for $\phi^\pm$ resulting in a mass splitting $\Delta M_\phi=M_{\phi^\pm}-M_{\phi^0}$.
The calculation of $\Delta M_\phi$ in our scenario is the same as the calculation of electromagnetic mass difference of pseudogoldstone
$\pi$-mesons, $\Delta M_\pi=M_{\pi^\pm}-M_{\pi^0}$. The one-loop result for the latter is well known,
\be
\label{13}
M_{\pi^\pm}^2-M_{\pi^0}^2=\frac{3\alpha}{8\pi F_\pi^2}\int_0^\infty dQ^2Q^2\left[\Pi_A(Q^2)-\Pi_V(Q^2)\right],
\ee
where $\Pi_V$ and $\Pi_A$ are the vector and axial correlators defined as in~\eqref{6}.
The result~\eqref{13} was first derived in 1967~\cite{Das} using the current algebra techniques. The modern derivation is based on the
method of effective action. The calculation of the corresponding Coleman-Weinberg potential leading to~\eqref{13} is nicely
reviewed in~\cite{contino}. Importantly, this derivation shows that the relation~\eqref{13} represents actually a particular case of a more
general result: The one-loop radiative correction to the mass of charged Goldstone bosons is proportional to
$\int dQ^2Q^2(\Pi_\text{br}-\Pi_\text{unbr})$, where $\Pi_\text{br}$ and $\Pi_\text{unbr}$ are the two-point correlators of currents
corresponding to broken and unbroken generators of spontaneously broken global symmetry. This is exploited, in particular, in the
$SO(5)/SO(4)$ scenario of the composite Nambu-Goldstone Higgs boson to generate the Higgs mass via radiative corrections from
hypothetical BSM strong sector (a pedagogical review is given in Ref.~\cite{contino}).

Using~\eqref{7}--\eqref{9} with the replacements mentioned after~\eqref{12}, one arrives at the relation by Das et al.~\cite{Das},
\be
\label{14}
M_{\pi^\pm}^2-M_{\pi^0}^2=\frac{3\alpha}{4\pi}\frac{M_{a_1}^2M_\rho^2}{M_{a_1}^2-M_\rho^2}\log\left(\frac{M_{a_1}^2}{M_\rho^2}\right).
\ee
It should be emphasized that the convergence in~\eqref{13} is provided by the asymptotic behavior~\eqref{7} for
$\left(\Pi_A(Q^2)-\Pi_V(Q^2)\right)$. The positivity of~\eqref{14} follows from the fact that the radiative corrections align the
vacuum along the direction preserving the $U(1)$ gauge symmetry, i.e., $\langle\pi^+\rangle=\langle\pi^-\rangle=0$ in the minimized
pion potential, so that the photon remains massless.

It is important to note that the relation~\eqref{14} was derived in the limit of massless pions. When the quark masses are turned on,
both $\pi^\pm$ and $\pi^0$ get a mass becoming pseudogoldstone bosons. The difference $\Delta M_\pi=M_{\pi^\pm}-M_{\pi^0}$, however,
remains dominated by the electromagnetic correction. This means that the electromagnetic pion mass difference~\eqref{14} arises at distances much
smaller than the scale of spontaneous CSB in QCD, $r_\text{\tiny CSB}\simeq 0.2$~fm. At distances
$r\ll r_\text{\tiny CSB}$ the pion can be considered as effectively massless. Assuming $M_{\pi^\pm}-M_{\pi^0}\ll M_\pi$, where
$M_\pi=M_{\pi^\pm}$ or $M_\pi=M_{\pi^0}$, we can write $M_{\pi^\pm}^2-M_{\pi^0}^2\simeq2M_\pi\Delta M_\pi$ and get the observable
value of $\Delta M_\pi$ substituting into
\be
\label{15}
\Delta M_\pi\simeq\frac{3\alpha}{8\pi}\frac{M_{a_1}^2M_\rho^2}{M_\pi(M_{a_1}^2-M_\rho^2)}\log\left(\frac{M_{a_1}^2}{M_\rho^2}\right),
\ee
the observable values of meson masses measured at larger distances, where the meson masses arise from a confinement mechanism.
Essentially the same trick we are going to use for calculation of $\Delta M_W=M_{W^\pm}-M_{W^0}$.

Under our assumptions, we are ready now to write the answer for $M_{\phi^\pm}^2-M_{\phi^0}^2$ directly from~\eqref{14},
\be
\label{16}
M_{\phi^\pm}^2-M_{\phi^0}^2=\frac{3\alpha}{4\pi}\frac{M_{W'}^2M_W^2}{M_{W'}^2-M_W^2}\log\left(\frac{M_{W'}^2}{M_W^2}\right).
\ee
It should be noted that, if our assumptions are true, the relation~\eqref{16} can turn out to be much more precise than~\eqref{14}.
Indeed, the relation~\eqref{14} was derived using two rough approximations --- infinitely narrow decay width and neglecting
contributions of radial excitations. The real $\rho$ and $a_1$ mesons, however, are broad resonances for which the
ratio $\Gamma/M$ is not small: $\Gamma_\rho\approx150$~MeV, $M_\rho\approx775$~MeV, $\Gamma_{a_1}\approx420$~MeV, $M_{a_1}\approx1230$~MeV~\cite{pdg}.
Quite surprisingly, the theoretical prediction from~\eqref{15}, $\Delta M_\pi^\text{\tiny (th)}\approx5.8$~MeV, agrees reasonably with
the experimentally measured value, $\Delta M_\pi^\text{\tiny (exp)}\approx4.6$~MeV~\cite{pdg}. Concerning the second approximation, the $\rho$ and
$a_1$ mesons, as all hadrons, are composite systems of quarks bound by strong interactions and this leads to the existence of towers
of radially excited $\rho$ and $a_1$ mesons which are listed in the Particle Data~\cite{pdg}. These excited states contribute to the $\rho$ and $a_1$
analogues of correlators~\eqref{8} and~\eqref{9} via additional pole terms. The resulting modification of~\eqref{14} seems to improve
the quantitative agreement between $\Delta M_\pi^\text{\tiny (th)}$ and $\Delta M_\pi^\text{\tiny (exp)}$~\cite{Yad}.
In the case under consideration, the ratio $\Gamma_W/M_W$ is smaller by an order of magnitude and the $W$-boson, as a true elementary particle,
does not have radial excitations.

Let us now motivate why we expect the fulfillment of the relation
\be
\label{17}
M_{W^\pm}^2-M_{W^0}^2\simeq M_{\phi^\pm}^2-M_{\phi^0}^2.
\ee
On the scales where the standard Higgs mechanism starts to work, $\phi^\pm$ and $\phi^0$ become the longitudinal components of
$W^\pm$ and $W^0$ gauge bosons. The $W^\pm$-bosons produced in the CDF experiment at Tevatron are ultrarelativistic\footnote{They were
produced in proton-antiproton collisions at a center of mass energy $E_\text{c.m.}=1.96$~TeV~\cite{CDF}. One can estimate the average proper
energy of each produced $W$-boson as $E_W\simeq\frac13\cdot\frac12\cdot E_\text{c.m.}\approx4M_W$, where the factor of $\frac13$ takes into account
that only one of three available quark-antiquark pairs produces the $W$-boson and $\frac12$ emerges from the well known experimental fact
that the quark degrees of freedom carry about a half of momentum of ultrarelativistic nucleon.}. It is easy to show that the longitudinal
polarization $\epsilon_\text{L}^\mu$ of such a $W$-boson becomes increasingly parallel to its four-momentum $k^\mu=(E_W,0,0,k)$ as $k$
becomes large (see, e.g., the classical textbook~\cite{Peskin}),
\be
\label{18}
\epsilon_\text{L}^\mu(k)=\frac{k^\mu}{M_W}+\mathcal{O}\left(\frac{M_W}{E_W}\right),\qquad k\to\infty.
\ee
Since the transverse polarizations $\epsilon_\bot^\mu$ do not grow with $k$, one can show that the physics of ultrarelativistic $W$-boson is
almost completely determined by its component $\epsilon_\text{L}^\mu$: The amplitude for emission or absorption of such $W$-bosons becomes
equal, at high energy, to the amplitude of emission or absorption of its longitudinal component. This statement constitutes the
essence of important {\it Goldstone boson equivalence theorem}: A relativistically moving, longitudinally polarized massive gauge boson behaves
as a Goldstone boson that was eaten by the Higgs mechanism~\cite{Peskin}. Since the mass of ultrarelativistic $W$-boson is also mostly determined
by its longitudinal component $\phi$, we should expect the relation~\eqref{17}.

Combining~\eqref{16} and~\eqref{17} we get the expression for $M_{W^\pm}^2-M_{W^0}^2$. Since
$\Delta M_W=M_{W^\pm}-M_{W^0}\ll M_W$ one can write $M_{W^\pm}^2-M_{W^0}^2\simeq2M_W\Delta M_W$. The result
for $\Delta M_W$ is
\be
\label{19}
\Delta M_W\simeq\frac{3\alpha}{8\pi}\frac{M_W M_{W'}^2}{M_{W'}^2-M_W^2}\log\left(\frac{M_{W'}^2}{M_W^2}\right).
\ee
As in the case of the pion analogue~\eqref{15}, the relation~\eqref{19} is derived below the scale
$r_\text{\tiny CSB}^\text{\tiny (weak)}$ where all particles are effectively massless. But the
observable value of $\Delta M_W$ at larger distances follows after substitution to~\eqref{19} the values
of $M_W$ and $M_{W'}$ at larger distances, where they emerge due to the Higgs mechanism.

Formally, the relation~\eqref{19} contains only one unknown parameter $M_{W'}$.
Another three unknown parameters $F_{W}$, $F_{W'}$ and $F_\phi$ are canceled due to the sum rules~\eqref{12}.
Following our suggestion, the last step is to take the degeneracy limit $M_{W'}=M_W$ since $W'$ represents
actually the same physical degree of freedom as $W$.
Using the limit $\frac{\log x}{x-1}\to1$ as $x\to1$
we obtain from~\eqref{19} our final result
\be
\label{20}
\Delta M_W\simeq\frac{3\alpha}{8\pi}M_W.
\ee
Substituting the experimental mass of $W$-boson, the relation~\eqref{20} predicts $\Delta M_W\simeq70.0$~MeV.
The given value is in perfect agreement with the observed discrepancy~\eqref{3}.

The physical meaning of additional $SU(2)'$ global symmetry above the electroweak scale is an open question.
To answer this question one should elaborate some other observable consequences of this symmetry.
We leave this for the future.

\bigskip

\noindent
{\it{Acknowledgements}}.
This research was funded by the Russian Science Foundation grant number 21-12-00020.


\begin{thebibliography}{99}

\bibitem{CDF}
T.~Aaltonen \textit{et al.} [CDF Collaboration],
``High-precision measurement of the W boson mass with the CDF II detector,''
Science \textbf{376} (2022) no.6589, 170-176.

\bibitem{pdg} P.A. Zyla {\it et al.} [Particle Data Group], ``Review of Particle Physics,''
Prog. Theor. Exp. Phys. {\bf 2020} (2020) no.8, 083C01.

\bibitem{deBlas:2022hdk}
J.~de Blas, M.~Pierini, L.~Reina and L.~Silvestrini,
``Impact of the recent measurements of the top-quark and W-boson masses on electroweak precision fits,''
arXiv:2204.04204 [hep-ph].

\bibitem{Lu:2022bgw}
C.~T.~Lu, L.~Wu, Y.~Wu and B.~Zhu,
``Electroweak Precision Fit and New Physics in light of $W$ Boson Mass,''
arXiv:2204.03796 [hep-ph].

\bibitem{Bhaskar:2022vgk}
A.~Bhaskar, A.~A.~Madathil, T.~Mandal and S.~Mitra,
``Combined explanation of $W$-mass, muon $g-2$, $R_{K^{(*)}}$ and $R_{D^{(*)}}$ anomalies in a singlet-triplet scalar leptoquark model,''
arXiv:2204.09031 [hep-ph].
\bibitem{raby}
J. Kawamura and S. Raby, ``$W$ mass in a model with vector-like leptons and $U(1)'$,''
arXiv:2205.10480 [hep-ph].
\bibitem{Dcruz:2022dao}
R.~Dcruz and A.~Thapa,
``$W$ boson mass, dark matter and $(g-2)_\ell$ in ScotoZee neutrino mass model,''
arXiv:2205.02217 [hep-ph].
\bibitem{Appelquist:2022qgl}
T.~Appelquist, J.~Ingoldby and M.~Piai,
``Composite two-Higgs doublet model from dilaton effective field theory,''
arXiv:2205.03320 [hep-ph].
\bibitem{Evans:2022dgq}
J.~L.~Evans, T.~T.~Yanagida and N.~Yokozaki,
``W boson mass anomaly and grand unification,''
arXiv:2205.03877 [hep-ph].

\bibitem{svz}
M.~A.~Shifman, A.~I.~Vainshtein and V.~I.~Zakharov,
``QCD and Resonance Physics. Theoretical Foundations,''
Nucl. Phys. B \textbf{147} (1979), 385-447.

\bibitem{Wein}
S.~Weinberg, ``Precise relations between the spectra of vector and axial vector mesons,''
Phys. Rev. Lett. \textbf{18} (1967), 507-509.

\bibitem{Das}
T.~Das, G.~S.~Guralnik, V.~S.~Mathur, F.~E.~Low and J.~E.~Young,
``Electromagnetic mass difference of pions,''
Phys. Rev. Lett. \textbf{18} (1967), 759-761.

\bibitem{contino}
R.~Contino, ``The Higgs as a Composite Nambu-Goldstone Boson,''
arXiv:1005.4269 [hep-ph].

\bibitem{Yad}
V.~A.~Andrianov and S.~S.~Afonin,
``Contribution of higher meson resonances to the electromagnetic $\pi$-meson mass difference,''
Phys. Atom. Nucl. \textbf{65} (2002), 1862-1867
[arXiv:hep-ph/0109026 [hep-ph]].

\bibitem{Peskin}
M.~E.~Peskin and D.~V.~Schroeder,
``An Introduction to quantum field theory,''
Addison-Wesley Publishing Company, 1995.

\end{thebibliography}
\end{document}